\documentclass[5p]{elsarticle}

\usepackage[T1]{fontenc}
\usepackage[utf8]{inputenc}
\usepackage{lmodern}        
\usepackage{textcomp}      
\usepackage{microtype}     
\usepackage{amsmath}
\usepackage{siunitx} 
\usepackage[version=4]{mhchem} 
\usepackage{graphicx}
\graphicspath{{./figures/}}
\usepackage{booktabs}       
\usepackage{multirow}      
\usepackage{caption}
\usepackage{float}

\makeatletter
\g@addto@macro\@floatboxreset\centering
\makeatother

\usepackage{hyperref}
\usepackage{breakurl}

\hypersetup{
    colorlinks=true,
    linkcolor={blue!80!black},
    citecolor={blue!80!black},
    urlcolor={blue!80!black}
}

\biboptions{sort&compress}

\begin{document}
\begin{sloppypar}

\title{
Interplay between crystal structure and magnetism in \ce{CeCrB4}
}

\author{Mirosław Werwiński\corref{cor1}}
\ead{werwinski@ifmpan.poznan.pl}

\author{Andrzej Szajek}
\author{Andrzej Kowalczyk}

\cortext[cor1]{Corresponding author.}

\address{Institute of Molecular Physics, Polish Academy of Sciences, M. Smoluchowskiego 17, 60-179 Poznań, Poland}

\begin{abstract}
\ce{CeCrB4} belongs to the family of \ce{MTB4} ternary metal borides, some of whose magnetic phases have recently been identified as quantum spin dimers.
Employing density functional theory within the GGA+$U$ framework, this work investigates the key ground-state properties of this system, including its ferromagnetic ground state, the mixed-valence nature of cerium, and the charge transfer from the cationic cerium--chromium layers to the anionic boron layers.
In particular, we focus on elucidating the bonding mechanism within the structurally embedded Cr--Cr dimers.
Specifically, we evaluate a previously proposed interpretation that the localized Cr~$3d_{z^2}$ states are molecular-like bonding and antibonding states.
The calculated dependence of the $3d_{z^2}$ electronic structure on the intra-dimer distance correlates with the level-splitting characteristics known for classical homonuclear diatomic molecules.
\end{abstract}

\maketitle

\section{Introduction}

\ce{MTB4} materials have recently attracted significant interest due to their structural motif of transition-metal dimers, which gives rise to the unique electronic and magnetic properties~\cite{sun_prediction_2023,zhang_unveiling_2024,zhang_high-throughput_2025,anas_investigation_2025}.
The \ce{M} site in the \ce{MTB4} formula can be occupied by metal cations, including \ce{Sc}, \ce{Y}, \ce{Mg}, \ce{Ca}, \ce{Al}, lanthanides, and actinides (\ce{U}, \ce{Th}).
Some \ce{MTB4} compounds containing Cr or Mn as the transition metal (T) have been classified as quantum spin-dimer magnets, exhibiting a magnetic singlet state of the transition-metal dimers~\cite{zhang_high-throughput_2025}.
This makes them, for example, promising candidates for studying the Bose-Einstein condensation of magnetic excitations in bulk~\cite{zapf_bose-einstein_2014, zhang_unveiling_2024}.
Moreover, because the localized electronic states of dimers depend strongly on their specific chemical composition (i.e., the selection of M and T), their properties can be precisely tailored.
Consequently, \ce{MTB4} materials may find novel applications in modern electronics and spintronics.

An interesting example within the \ce{MTB4} family is \ce{CeCrB4}, a stable compound that potentially contains two magnetic elements.
Although first synthesized in the 1970s~\cite{kuzma_sistemy_1973,burzo_rare_2023}, its magnetic properties have only been discussed generally within broader material screenings in two recent studies~\cite{flipo_thermoelectricity_2021,zhang_unveiling_2024}.
In contrast to the aforementioned quantum spin-dimer magnets, DFT calculations classify the ground state of \ce{CeCrB4} as a conventional antiferromagnet featuring a magnetic triplet state of Cr--Cr dimers~\cite{zhang_unveiling_2024}.
Furthermore, the same calculations revealed pairs of localized Cr~$3d$ states near the Fermi level for \ce{CeCrB4}, which were interpreted as related to bonding and antibonding molecular-type orbitals of the Cr--Cr dimers~\cite{flipo_thermoelectricity_2021,zhang_unveiling_2024} -- a feature that is also characteristic of other \ce{MTB4} compounds~\cite{sun_prediction_2023,zhang_high-throughput_2025,anas_investigation_2025}.
However, the calculated difference in total energy between the antiferromagnetic and ferromagnetic states was minor ($\sim$0.3~meV\,atom$^{-1}$), falling within the accuracy limit of the method applied~\cite{zhang_unveiling_2024}.
Therefore, it is worthwhile to reinvestigate the magnetic ground state of \ce{CeCrB4} using a different DFT code while ensuring the highest possible numerical precision.
Furthermore, although interpreting the localized $3d$ state pairs as molecular-type bonding and antibonding states appears plausible, no detailed studies have yet been conducted on any material within the \ce{MTB4} family to definitively verify this hypothesis.

The purpose of this work is to clarify the molecular nature of the localized dimer states in \ce{MTB4} compounds, determine the magnetic ground state of \ce{CeCrB4}, and resolve the valence state and magnetic moment of Ce in this material.
Verifying the molecular nature of dimer states will reinforce the significance of \ce{MTB4} systems for currently explored applications, while also drawing interest from related disciplines, such as surface science~\cite{zhang_electronic_2025}, thin films, high entropy alloys~\cite{yang_synthesis_2023}, and systems with magnetic anisotropy.
Furthermore, dimers with molecular-type bondings should exhibit a strong dependence of their localized state energies on the intra-dimer distance.
Controlling this distance via external pressure or, in the case of thin films, through substrate-induced lattice strain provides an additional degree of freedom to manipulate the electronic structure near the Fermi level. 
This, in turn, may trigger significant modifications in the material properties, including electronic and magnetic phase transitions.

Here, we elucidate whether the transition-metal dimers in \ce{MTB4} compounds possess a molecular character.
To achieve this, we perform DFT calculations for a nonmagnetic \ce{CeCrB4} model, analyzing the evolution of the electronic density of states (DOS) as a function of the Cr--Cr dimer distance.
We also compute precise total energies for both the ferromagnetic and antiferromagnetic configurations, incorporating optimization of the atomic positions, to identify the magnetic ground state.

%
\section{\label{sec:comp_details} Computational details}

\begin{figure}[!ht]
    \centering
    \includegraphics[trim = 0 0 0 0, clip,width=0.95\columnwidth]{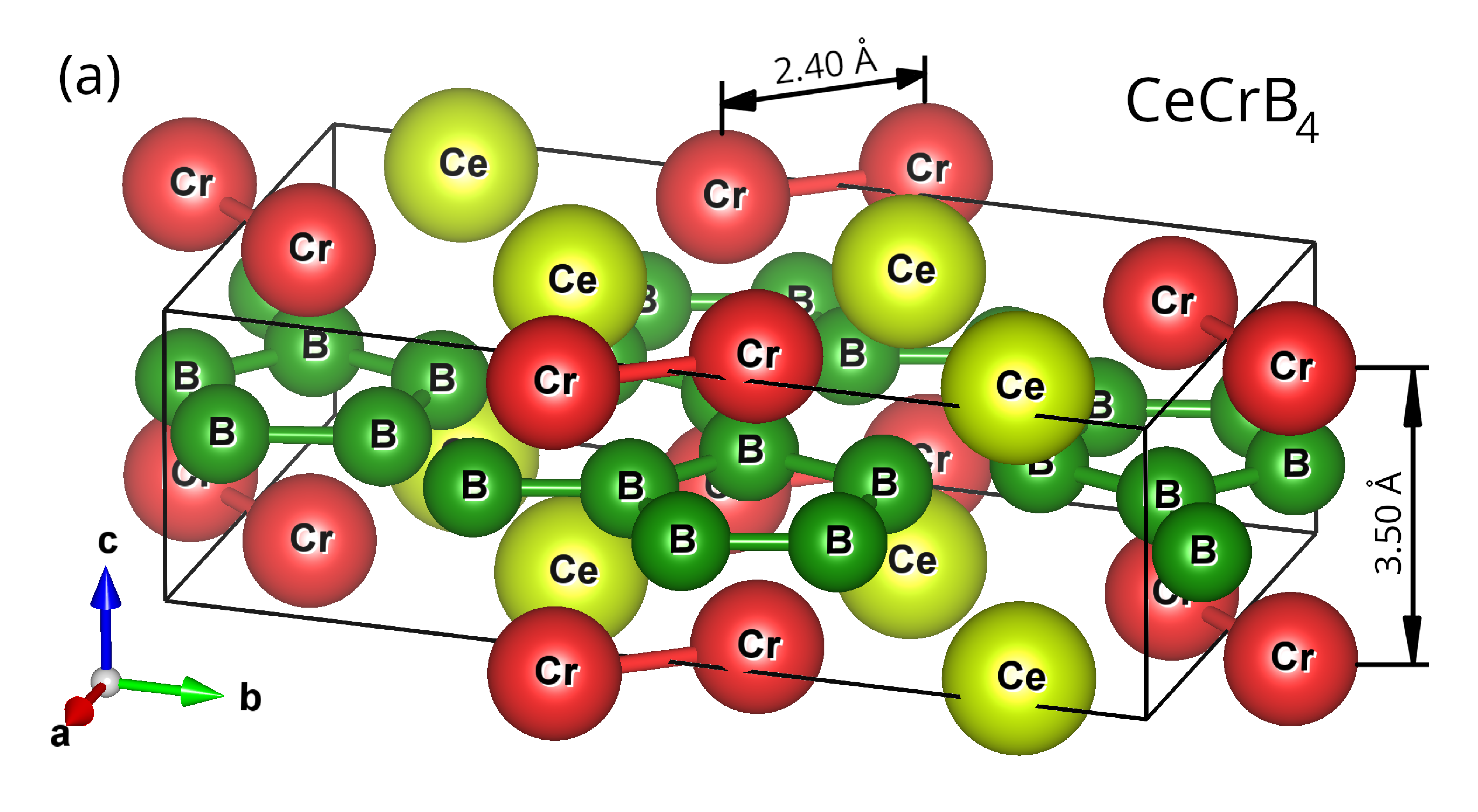}
    \includegraphics[trim = 0 0 0 0, clip,width=0.85\columnwidth]{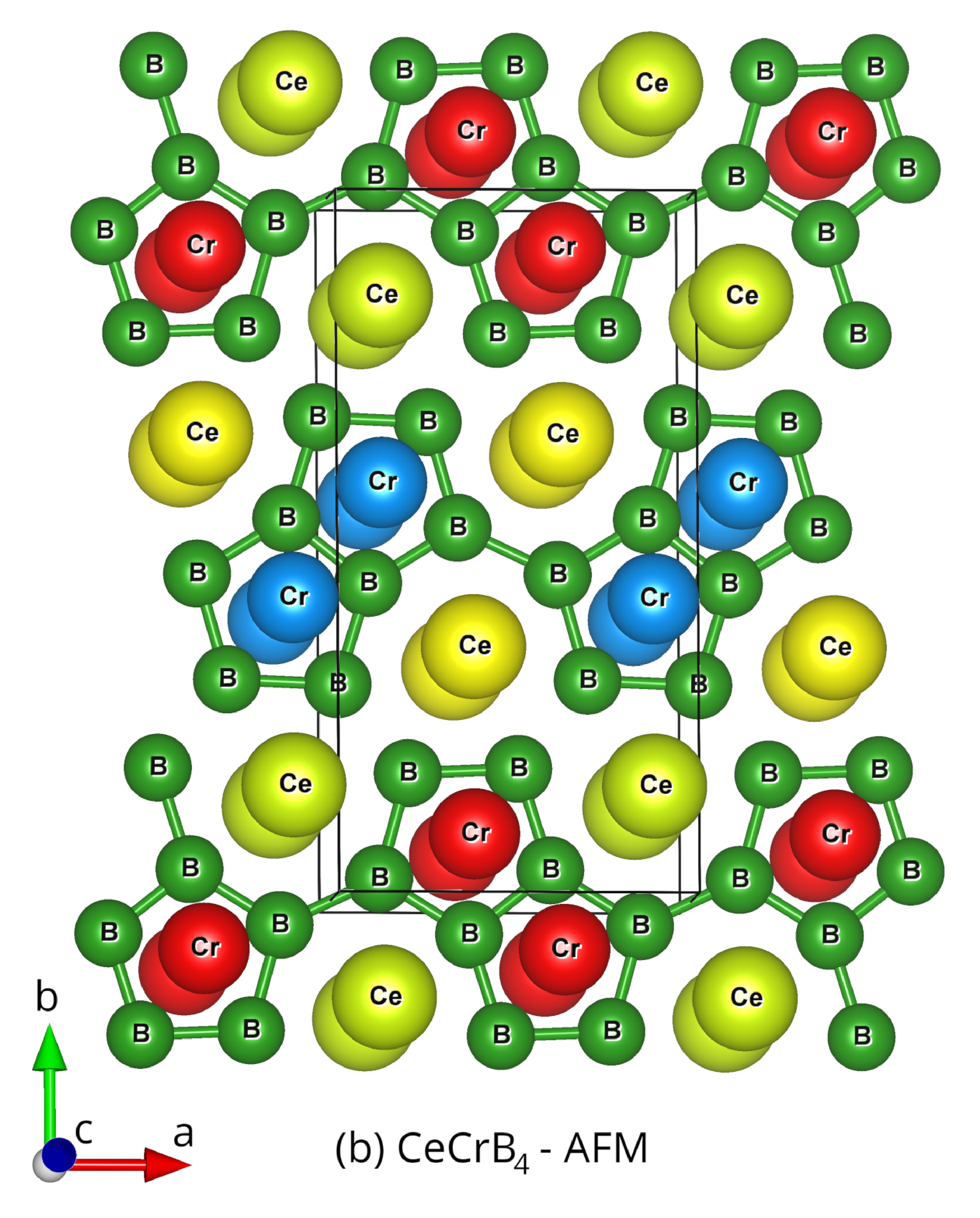}
    \caption{
\ce{CeCrB4} crystallizes in the orthorhombic \ce{YCrB4}-type structure (space group $Pbam$) with lattice parameters $a = 5.942$~\AA{}, $b = 11.580$~\AA{}, and $c = 3.496$~\AA{} (our experimental data).
(a) Single unit cell with the intra-dimer Cr--Cr distances indicated.
(b) Antiferromagnetic configuration featuring two magnetically inequivalent Cr sites, highlighted in different colors.
The extended projection help to visualise the Cr--Cr dimers within the lattice.
}
    \label{fig:crystal_structure}
\end{figure}

\begin{table}[h]
\centering
\caption{
Crystallographic data for \ce{CeCrB4}.
Experimental lattice parameters ($a = 5.942$~\AA{}, $b = 11.580$~\AA{}, $c = 3.496$~\AA{}) and Wyckoff positions optimized via DFT. 
The left and right columns correspond to the ferromagnetic (FM, space group $Pbam$) and antiferromagnetic (AFM, space group $P2/m$, No.~10, unique axis $a$ setting) configurations.
The optimization of atomic positions was performed using the spin-polarized  scalar-relativistic FPLO method within the PBE+$U$ approach ($U_{4f} = 6$~eV).
}
\label{tab:crystallographic_data}
\small 
\setlength{\tabcolsep}{4pt} 
\begin{tabular}{c *{2}{S[table-format=1.4]} S[table-format=1.1] c  c *{2}{S[table-format=1.4]} S[table-format=1.1]}
\hline \hline
\multicolumn{4}{c}{FM model} & & \multicolumn{4}{c}{AFM model} \\
\cline{1-5} \cline{6-9}
  & {\textbf{$x$}} & {\textbf{$y$}} & {\textbf{$z$}} & &   & {\textbf{$x$}} & {\textbf{$y$}} & {\textbf{$z$}} \\
\hline 
Ce & 0.1237 & 0.1512 & 0.0 & & Ce & 0.8764 & 0.1511 & 0.0 \\
Cr & 0.1240 & 0.4177 & 0.0 & & Ce & 0.6236 & 0.6511 & 0.0 \\
B  & 0.2799 & 0.3164 & 0.5 & & Cr & 0.8759 & 0.4178 & 0.0 \\
B  & 0.3615 & 0.4668 & 0.5 & & Cr & 0.6241 & 0.9178 & 0.0 \\
B  & 0.3870 & 0.0459 & 0.5 & & B  & 0.7198 & 0.3164 & 0.5 \\
B  & 0.4738 & 0.1902 & 0.5 & & B  & 0.7802 & 0.8164 & 0.5 \\
   &        &        &     & & B  & 0.6385 & 0.4667 & 0.5 \\
   &        &        &     & & B  & 0.8615 & 0.9667 & 0.5 \\
   &        &        &     & & B  & 0.6130 & 0.0459 & 0.5 \\
   &        &        &     & & B  & 0.8870 & 0.5459 & 0.5 \\
   &        &        &     & & B  & 0.5263 & 0.1902 & 0.5 \\
   &        &        &     & & B  & 0.9737 & 0.6902 & 0.5 \\
\hline \hline 
\end{tabular}
\end{table}

%
For calculations, we used density functional theory (DFT) as implemented in the full-potential local-orbital (FPLO18.00-52) code~\cite{koepernik_full-potential_1999}.
To account for spin-orbit coupling, we applied a fully relativistic approach~\cite{eschrig_chapter_2004}.
For the exchange-correlation potential, we selected the generalized gradient approximation (GGA) in the Perdew-Burke-Ernzerhof (PBE) parameterization~\cite{perdew_generalized_1996}.
As suggested by Zhang~\textit{et~al.}~\cite{zhang_unveiling_2024}, to improve the description of Ce~4$f$ orbitals in \ce{CeCrB4}, we introduced an additional Coulomb~($U$) term accounting for intra-atomic repulsion into the energy functional~\cite{ylvisaker_anisotropy_2009}.
We employed the fully localized limit (FLL) of the GGA+$U$~\cite{czyzyk_local-density_1994}, which is also referred to as the atomic limit.
We have set the value of the effective Coulomb $U$ parameter for the Ce~4$f$ orbitals to 6~eV, which is common value for Ce compounds with localized $f$ electrons~\cite{rusz_probing_2008,winiarski_electronic_2013}.

%
For the Brillouin zone integration, we used a $9 \times 7 \times 15$~$k$-mesh.
The self-consistency was achieved with a total energy convergence condition of $10^{-8}$~Ha ($2.72 \times 10^{-7}$~eV).
The atomic positions were optimized with spin-polarization in sclar-relativistic approach using forces with a convergence criterion of $10^{-3}$~eV\,\AA$^{-1}$.
To visualize the crystal structure we have used VESTA~\cite{momma_vesta_2008}.

%
\ce{CeCrB4} crystallizes in an orthorhombic \ce{YCrB4}-type structure (space group $Pbam$)~\cite{kuzma_sistemy_1973,braun_smfeb4_1980, flipo_thermoelectricity_2021, burzo_rare_2023, tokuda_redetermination_2023, zhang_unveiling_2024}, see Fig.~\ref{fig:crystal_structure} and Table~\ref{tab:crystallographic_data}.
In the experimental work conducted in parallel by us, which is not discussed here, we determined the lattice parameters of \ce{CeCrB4} as $a = 5.942$~\AA{}, $b = 11.580$~\AA{}, $c = 3.496$~\AA{}, which we use along this paper.
They agree well with the corresponding values determined by Kuźma~\textit{et~al.}: $a = 5.974 \pm 0.005$~\AA{}, $b = 11.53 \pm 0.01$~\AA{}, $c = 3.536 \pm 0.004$~\AA{}~\cite{kuzma_sistemy_1973}.

The crystal structure of \ce{CeCrB4} is characterized by a layered arrangement.
Boron atoms form a planar lattice consisting of alternating 5- and 7-membered rings.
Chromium atoms form Cr--Cr dimers oriented in the $ab$ plane with a relatively small interatomic distance of $d_{\text{Cr-Cr}} = 2.40$~\AA{},
where the distance between the Cr atoms along the $c$-axis is much larger and equal to the lattice parameter of $3.50$~\AA{}, see Fig.~\ref{fig:crystal_structure}(a).

A previous study examining the \ce{MTB4} class analyzed a number of possible magnetic configurations for each compound~\cite{zhang_unveiling_2024}. 
For \ce{CeCrB4}, two magnetic configurations -- the ferromagnetic and antiferromagnetic (of the FAF type, according to the Zhang~\textit{et al.} nomenclature) -- had the lowest, nearly identical total energies ($\Delta E = 0.3$~\text{meV}\,\text{atom}$^{-1}$, with antiferromagnetic ground state)~\cite{zhang_unveiling_2024}.
We selected these two magnetic configurations, both characterized by the parallel alignment of magnetic moments within the Cr--Cr dimers, for our study, see Fig.~\ref{fig:crystal_structure}(1) and Table~\ref{tab:crystallographic_data}.

%
It should be noted, however, that the computational results presented here are inherently subject to the limitations of the DFT framework.
By definition, these calculations describe the ground state, which corresponds to a temperature of $0$~K.
Moreover, the choice of the GGA-PBE functional is known to affect the localization of electrons, which may influence the values of the calculated magnetic moments.
Furthermore, while the applied Coulomb correction improves the description of $\text{Ce}~4f$ orbitals, it cannot fully capture all dynamic many-body correlation effects.
Nevertheless, bearing these approximations in mind, our results provide robust insights into the interplay between the unique crystal structure of \ce{CeCrB4} and its electronic and magnetic properties.

%
\section{Results and discussion}

\begin{table}[h]
\centering
\caption{
Spin ($m_{\mathrm{s}}$) and orbital ($m_{\mathrm{l}}$) magnetic moments (in $\mu_{\mathrm{B}}$\,atom$^{-1}$) calculated for \ce{CeCrB4} in the ferromagnetic (FM) and antiferromagnetic (AFM) configurations. 
In the FM state, the total spin magnetic moment is $m_{\mathrm{s}}^{\mathrm{tot}} \approx 0.388~\mu_{\mathrm{B}}$\,f.u.$^{-1}$ and the total orbital magnetic moment is $m_{\mathrm{l}}^{\mathrm{tot}} \approx -0.035~\mu_{\mathrm{B}}$\,f.u.$^{-1}$.
The calculations were performed using the fully relativistic FPLO method within the PBE+$U$ approach ($U_{4f} = 6$~eV).
}
\label{tab:magnetic_moments}
\begin{tabular}{ c *{2}{S[table-format=-1.3] S[table-format=-1.3] }}
\hline  \hline
 & \multicolumn{2}{c}{FM state} & \multicolumn{2}{c}{AFM state}  \\
\cline{2-3} \cline{4-5} 
Element & {$m_{\mathrm{s}}$} & {$m_{\mathrm{l}}$} & {$m_{\mathrm{s}}$ } & {$m_{\mathrm{l}}$} \\
\hline
Ce &  0.027 & -0.020 & \pm 0.014 & \pm 0.003 \\
Cr &  0.401 & -0.014 & \pm 0.414 & \mp 0.012 \\
B  & -0.010 & -0.001 & \pm 0.012 &   0.000   \\
B  & -0.008 &  0.000 & \pm 0.009 &   0.000   \\
B  & -0.016 &  0.000 & \pm 0.017 &   0.000   \\
B  & -0.007 &  0.000 & \pm 0.010 &   0.000   \\
\hline \hline
\end{tabular}
\end{table}

\begin{table}[t]
\centering
\caption{
Mulliken population analysis for the valence states of \ce{CeCrB4} in the ferromagnetic state. 
The valence electron configurations and effective charges ($\Delta Q$) are given.
The calculations were performed using the fully relativistic FPLO method within the PBE+$U$ approach ($U_{4f} = 6$~eV).
}
\label{tab:mulliken}
\begin{tabular}{ c *{4}{l } r }
\hline \hline
El. & \textbf{$s$} & \textbf{$p$} & \textbf{$d$} & \textbf{$f$} & {$\Delta Q$} \\
\hline
Ce & 6$s$: 0.19 & 5$p$: 5.82 & 5$d$: 1.69 & 4$f$: 0.66 & +1.45 \\
Cr & 4$s$: 0.43 & 4$p$: 0.48 & 3$d$: 4.60 & --        & +0.54 \\
B  & 2$s$: 1.16 & 2$p$: 2.31 &  -- & --        & -0.53 \\
B  & 2$s$: 1.15 & 2$p$: 2.31 &  -- & --        & -0.51 \\
B  & 2$s$: 1.11 & 2$p$: 2.24 &  -- & --        & -0.41 \\
B  & 2$s$: 1.16 & 2$p$: 2.31 &  -- & --        & -0.53 \\
\hline \hline
\end{tabular}
\end{table}

%
A distinctive feature of \ce{MTB4} compounds is the presence in their crystal structure of transition metal dimers~\cite{blonski_magnetic_2009}, see Fig.~\ref{fig:crystal_structure}(a).
The interatomic distance within these dimers falls well within the range of covalent-bond length (e.g., 2.40~\AA{} for \ce{CeCrB4}).
Furthermore, dimers such as Cr--Cr or Mn--Mn form magnetic, molecular-like states: antiparallel ($\uparrow\downarrow$, singlet, $S = 0$) or parallel ($\uparrow\uparrow$, triplet, $S = 1$)~\cite{zhang_unveiling_2024,zhang_high-throughput_2025}.
Which of these configurations constitutes the ground state is dictated by the specific \ce{M} element~\cite{zhang_unveiling_2024,zhang_high-throughput_2025}.
In the case of \ce{CeCrB4}, DFT-GGA calculations indicated that the magnetic ground state of the Cr--Cr dimers is a triplet ($\uparrow\uparrow$)~\cite{zhang_unveiling_2024}.
These magnetic dimers, in turn, couple to form ferromagnetic or antiferromagnetic bulk phases~\cite{zhang_unveiling_2024}, see Fig.~\ref{fig:crystal_structure}(b).
In this work, we focus on the relationship between the magnetic properties of the compound and the Cr--Cr dimer motif in the \ce{CeCrB4} crystal structure.

%
The effective magnetic moment of \ce{CeCrB4}, determined from the modified Curie-Weiss law for $T > 100$~K, is 0.6~$\mu_\mathrm{B}$~\cite{flipo_thermoelectricity_2021}.
Flipo~\textit{et al.}~\cite{flipo_thermoelectricity_2021} attributed this value to the Ce$^{3+}$ ions.
However, Zhang~\textit{et al.} reported a magnetic moment of 0.35~$\mu_\mathrm{B}$ per Cr atom based on DFT-GGA calculations (without $U$) for the antiferromagnetic bulk phase hosting Cr--Cr triplet dimers~\cite{zhang_unveiling_2024}.
The magnetic susceptibility of \ce{CeCrB4} measured down to 1.8~K revealed no magnetic phase transition~\cite{flipo_thermoelectricity_2021}.

\subsection{Ferromagnetic ground state and magnetic moment of \mbox{Cr--Cr~dimers}}

%
In our study, we reinvestigated the two lowest-energy magnetic configurations of \ce{CeCrB4} proposed by Zhang~\textit{et al.}~\cite{zhang_unveiling_2024}.
Specifically, we focused on the ferromagnetic and antiferromagnetic (FAF~\cite{zhang_unveiling_2024}) structures, both hosting the triplet state of the Cr--Cr dimers.
The optimized atomic positions for both magnetic configurations are summarized in Table~\ref{tab:crystallographic_data}.
Our GGA+$U$~(Ce~4$f$) calculations indicate that the ferromagnetic state is energetically more favorable by 0.15~meV\,atom$^{-1}$.
Using the relation $E = k_{\text{B}}T$, this energy difference corresponds to a temperature of 1.72~K, which suggests why experimental measurements conducted down to 1.8~K failed to detect a magnetic phase transition.

%
The calculated magnetic moments for both the FM and AFM states are compiled in Table~\ref{tab:magnetic_moments}.
Regarding the individual sublattices, the results for both magnetic configurations are similar.
The Ce sublattice exhibits magnetic moments close to zero, whereas the Cr sublattice carries magnetic moments of approximately 0.40~$\mu_\mathrm{B}$.
This is consistent with the corresponding value of 0.35~$\mu_\mathrm{B}$ reported previously from GGA calculations~\cite{zhang_unveiling_2024}.

\begin{figure}
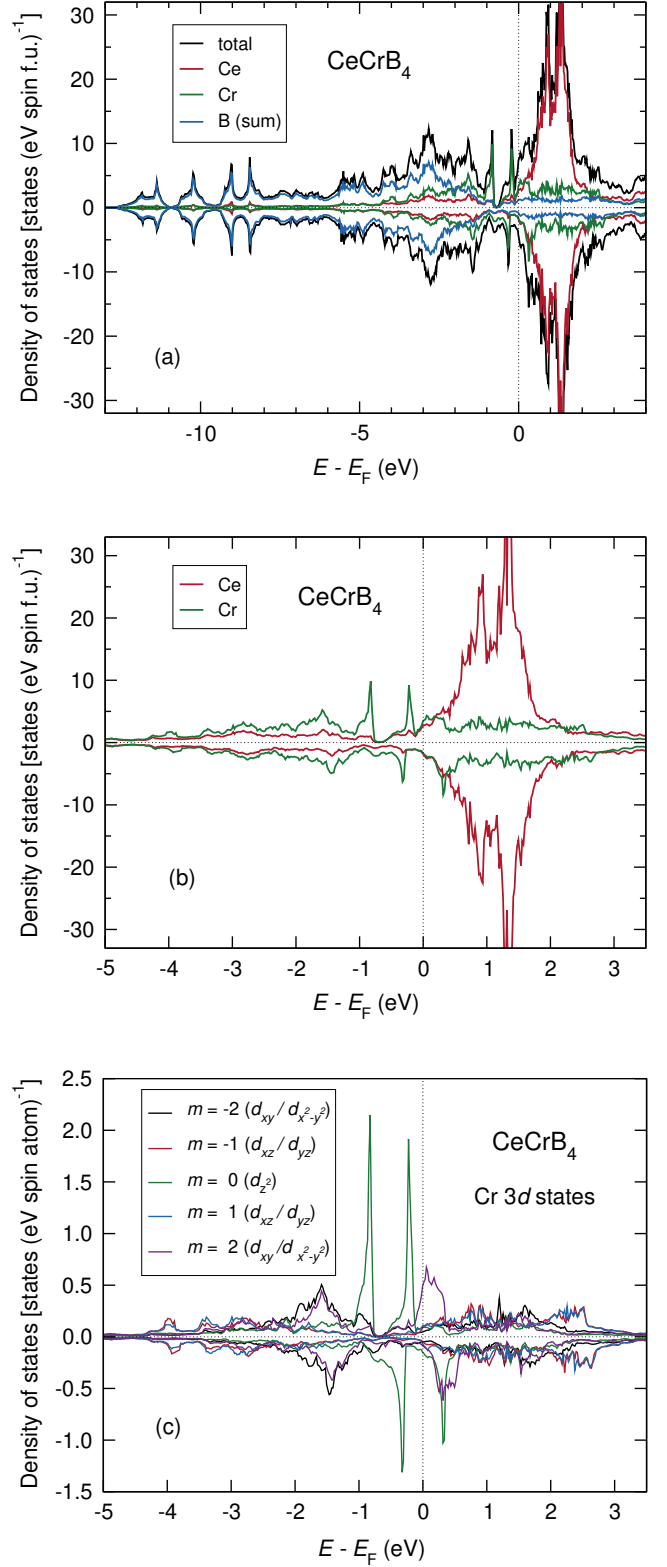

      \includegraphics[trim = 0 0 0 0, clip,width=\linewidth]{cecrb4_dos_pbe_uCe_6.eps} 
      \vspace{2mm}
      
      \includegraphics[trim = 0 0 0 0, clip,width=\linewidth]{cecrb4_dos_pbe_uCe_6_narrow.eps}
      \vspace{2mm}
      
      \includegraphics[trim = 0 0 0 0, clip,width=\linewidth]{cecrb4_dos_pbe_uCe_6_Cr_3d_local.eps}
    \caption{
Electronic density of states (DOS) for \ce{CeCrB4} in the ferromagnetic configuration:
(a)~total and site-projected DOS;
(b)~site-projected Ce and Cr DOS;
(c)~local Cr~$3d$ orbital contributions.
The results come from fully relativistic calculations with the magnetization (quantization) axis [001].
The calculations were performed using the FPLO method within the PBE+$U$ approach ($U_{4f} = 6$~eV).
}
    \label{fig:dos_cecrb4}
  \end{figure}

%
The magnetic moments on the Cr--Cr dimers originate predominantly from the $3d_{z^2}$ ($m = 0$) states. 
These are characterized by four strongly localized $3d_{z^2}$ states per dimer; two in each spin channel, see Fig.~\ref{fig:dos_cecrb4}(c).
If one of two states lie below the Fermi level, the magnetic moment of the Cr atom is $1~\mu_\mathrm{B}$ ($0~\mu_\mathrm{B}$ per dimer in singlet state).
If, on the other hand, two $3d_{z^2}$ states from one spin channel and one state from the opposite spin channel are occupied, the total magnetic moment of the dimer reduces to $1~\mu_\mathrm{B}$ ($0.5~\mu_\mathrm{B}$ per Cr atom) -- also our case, see Fig.~\ref{fig:dos_cecrb4}(c).

\begin{figure*}[t]

\centering
\includegraphics[trim = 0 0 0 0,clip,width=0.77\textwidth]{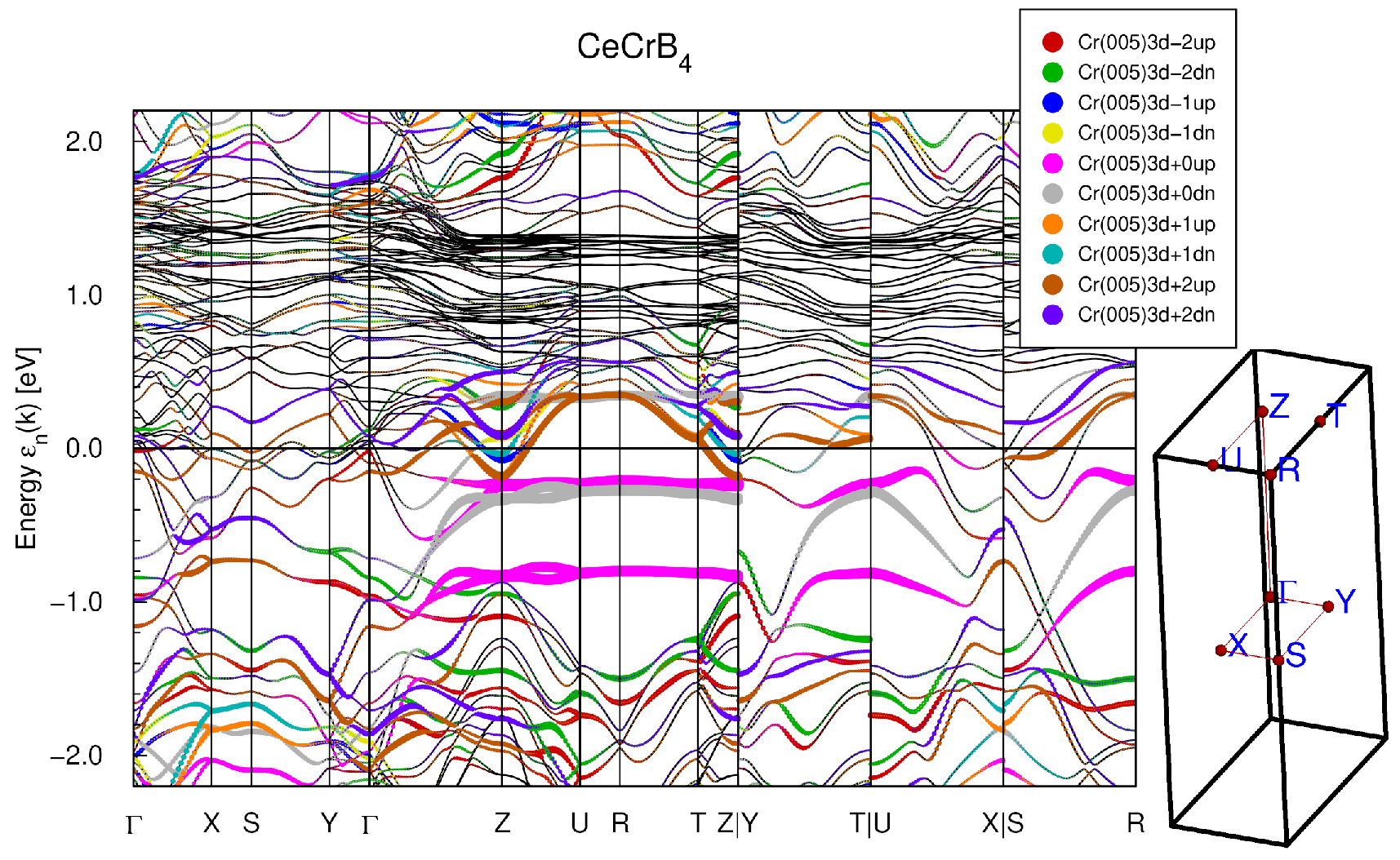}
\caption{   
Electronic band structure of \ce{CeCrB4} in the ferromagnetic configuration, presented in a fat-band representation.  
The local Cr~$3d$ orbital contributions are highlighted in color.
The partially flat Cr~$3d_{z^2}$ ($m = 0$) bands are indicated in magenta and gray for the two respective spin channels.
The corresponding high-symmetry points are marked on the enclosed Brillouin zone.
The calculations were performed using the fully relativistic FPLO method within the PBE+$U$ approach ($U_{4f} = 6$~eV).
\label{fig:bands}   
}
\end{figure*}

Zhang \textit{et al.}~\cite{zhang_unveiling_2024} demonstrated that magnetic compounds within the \ce{MTB4} family can be classified into two distinct categories: quantum spin dimers and conventional antiferromagnets, characterized by transition metal magnetic moments of approximately 1 and $0.5~\mu_\mathrm{B}$, respectively.
Consequently, antiferromagnetic \ce{CeCrB4}, with its Cr magnetic moments of about 0.35--0.40~$\mu_\mathrm{B}$, falls into the second category.
However, the ferromagnetic \ce{CeCrB4} does not fit directly into that dual classification.

\subsection{Mixed-valence state of Ce}

%
Structural investigations of the isostructural \ce{REReB4} series, where \ce{RE} stands for rare-earth element, revealed an anomaly in the lattice parameters for the Ce compound~\cite{abramchuk_crystal_2016}.
It indicates that the Ce configuration in \ce{CeReB4} deviates from the standard $4f^1$ ($\mathrm{Ce}^{3+}$) state toward either a mixed-valence $4f^1\text{--}4f^0$ or a pure $4f^0$ ($\mathrm{Ce}^{4+}$) configuration~\cite{abramchuk_crystal_2016}.

%
Accordingly, for \ce{CeCrB4}, high-energy resolution fluorescence-detected X-ray absorption spectroscopy (HERFD-XAS) indicates that Ce adopts a mixed-valence state of $3.54+$ at room temperature~\cite{flipo_thermoelectricity_2021}, which corresponds to an occupation of $0.46$ for the Ce~$4f$ states.
Furthermore, although the magnetic moment on Ce calculated from GGA+$U$ is close to zero (as shown earlier), the Ce~$4f$ shell is not empty, see Table~\ref{tab:mulliken}.
Specifically, the occupation of the Ce~$4f$ states yielded by the Mulliken population analysis~\cite{mulliken_electronic_1955} is $0.66$, corresponding to a nominal valence of approximately $3.34+$. 
It qualitatively confirms the mixed-valence nature of the Ce ions, despite underestimation compared to the mentioned experimental valence state ($3.54+$)~\cite{flipo_thermoelectricity_2021}.

Table~\ref{tab:mulliken} summarizes the occupations of the remaining valence orbitals alongside the effective charges at each crystallographic site.
The main electron donor in the system is Ce, carrying a substantial positive effective charge of $+1.45$.
It also exhibits strong intra-atomic hybridization, as indicated by the high occupancy of the $5d$ orbitals ($1.69$) at the expense of the $6s$ and $4f$ states.
Cr acts as a cation as well, carrying a charge of $+0.54$.
Together the Ce and Cr sublattices transfer approximately two electrons per formula unit to the borons sublattice.
From a structural perspective, this charge redistribution adds a distinct physical dimension to the layered character of crystal structure (composed of alternating Ce--Cr and B sheets), transforming it into a stack of alternately charged positive and negative layers.

\subsection{Van Hove singularities of Cr}

%
The magnetic and charge redistribution characteristics presented above are complemented by the calculated electronic density of states (DOS), see Fig.~\ref{fig:dos_cecrb4}.
The \ce{CeCrB4} compound exhibits distinct spin polarization.
Its metallic character, as evidenced by a finite DOS at the Fermi level, has already been inedified through resistivity measurements~\cite{flipo_thermoelectricity_2021}.
In the lower energy range from $-13$ to $-5$~eV, the densities of states are dominated by boron contributions.
Moving closer to the Fermi level, in the range from approximately $-5$ to $-1$~eV, a strong hybridization among the electronic states of all three constituent elements becomes evident.
In contrast, in the immediate vicinity of the Fermi level (from about $-1$~eV to little above $E_\mathrm{F}$), strongly localized Cr states stand out.
Finally, the relatively narrow and densely populated conduction band spanning from about $E_\mathrm{F}$ to $2$~eV is dominated by the Ce~$4f$ states.

The localized Cr states were initially identified in electronic structure of \ce{YCrB4} as Cr~$3d_{z^2}$ contributions (quantum number $m = 0$)~\cite{flipo_thermoelectricity_2021}.
Subsequently, they were interpreted as molecular-like states originating from the Cr--Cr dimers, where the bonding and antibonding states split into lower- and higher-energy states~\cite{zhang_unveiling_2024}.
Before evaluating this compelling hypothesis in the following subsection, we first analyze the local DOS contributions shown in Fig.~\ref{fig:dos_cecrb4}(c).
In contrast to previous studies, we present these electronic features with spin polarization.
Consequently, the two Cr~$3d_{z^2}$ peaks typically observed in the nonmagnetic state split within the ferromagnetic configuration into four distinct peaks.
Specifically, three of these states reside below the Fermi level, which accounts for the contribution to the magnetic moment of approximately 0.5~$\mu_\mathrm{B}$ per Cr atom.
Since localized electronic states in the vicinity of the Fermi level are known to possibly influence transport properties, these findings highlight the potential of \ce{MTB4}-type systems for tuning electrical conductivity via magnetic degrees of freedom.

Following the approach established in previous studies~\cite{flipo_thermoelectricity_2021,zhang_unveiling_2024}, we calculated the electronic band structure of the system, projected onto the local Cr~$3d$ orbital contributions, see Fig.~\ref{fig:bands}.
The strong localization of the $3d_{z^2}$ states, manifested by flat bands, occurs within the $Z\text{--}U\text{--}R\text{--}T$ plane of the Brillouin zone.
In contrast, this feature is absent in the $\Gamma\text{--}X\text{--}S\text{--}Y$ plane. 
This behavior can be intuitively understood by inspecting the unit cell, where the Cr atoms reside within the outer planes, see Fig.~\ref{fig:crystal_structure}(a).
Furthermore, the $3d_{z^2}$ states exhibit a pronounced dispersion along the $k_z$ direction (i.e., along the $\Gamma\text{--}Z$, $Y\text{--}T$, $S\text{--}R$, and $X\text{--}U$ paths), which stems from the out-of-plane hybridization of these orbitals with the 2$p$ states of boron atoms located in the adjacent layers.

\begin{figure}[t]
    \centering
    \includegraphics[trim = 40 0 0 0,clip,width=\columnwidth]{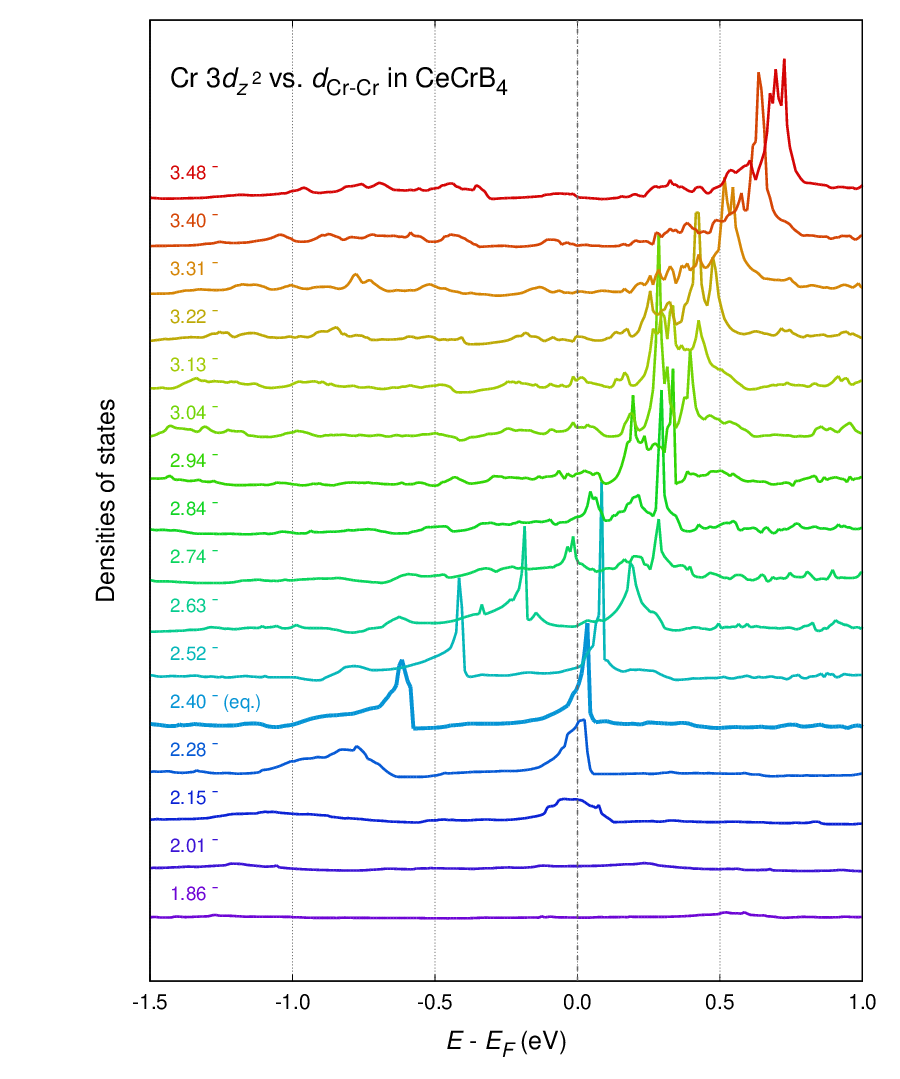}
    \caption{
Evolution of the orbital-resolved density of states (DOS) for the $\text{Cr}~3d_{z^2}$ ($m = 0$) state as a function of the intra-dimer $d_{\text{Cr--Cr}}$ distance in \ce{CeCrB4}. 
Individual spectra are vertically shifted by a constant offset for clarity. 
The vertical dashed line denotes the Fermi level ($E - E_{\mathrm{F}} = 0$~eV). 
The spectrum highlighted with a thicker line at $2.40$~\AA{} corresponds to the equilibrium configuration (eq.).
The calculations were performed using the nonmagnetic scalar-relativistic FPLO method within the PBE+$U$ framework ($U_{4f} = 6$~eV).
The variation in the interatomic distance $d_{\text{Cr--Cr}}$ was modeled by scaling the area of the $ab$ plane of the unit cell while keeping the lattice parameter $c$ constant.
}
\label{fig:3dz2_dos}
\end{figure}

  \begin{figure}[t]
  \centering
      \includegraphics[width=0.95\linewidth]{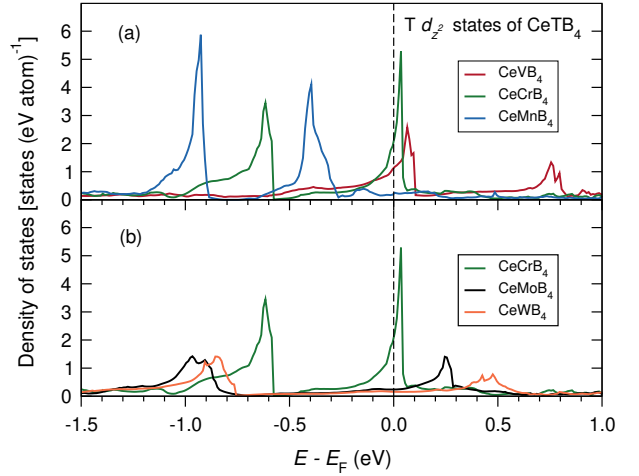}
    \caption{
Electronic density of states (DOS) for \ce{CeTB4} compounds:
(a)~local \ce{T}~$3d_{z^2}$ orbital contributions for \ce{T}~=~\ce{V, Cr}, and \ce{Mn} (consecutive elements within the $3d$ series);
(b)~local $3d_{z^2}$, $4d_{z^2}$, and $5d_{z^2}$ orbital contributions for \ce{T}~=~\ce{Cr, Mo}, and \ce{W}, respectively (three elements belonging to the same group of the periodic table).
The calculations were performed using the nonmagnetic scalar-relativistic FPLO method within the PBE+$U$ approach ($U_{4f} = 6$~eV).
For all considered compounds, the \ce{CeCrB4} structural model was utilized as a template, substituting only the chemical element at the Cr position.
}
    \label{fig:local_dos_cetb4}
  \end{figure}

\subsection{Tuning of Cr--Cr molecular-like states via interatomic distance}

%
Zhang~\textit{et al.} interpreted the pairs of localized Cr~$3d_{z^2}$ states as bonding and antibonding states originating from the Cr--Cr dimers~\cite{zhang_unveiling_2024}.
However, this compelling proposition has not yet been directly verified.
To test this hypothesis, we investigated the effect of varying the intra-dimer interatomic distance on the electronic structure of Cr~$3d_{z^2}$ states.

%
In classical homonuclear diatomic molecules, such as \ce{H2}, the energy levels of the bonding and antibonding states shift characteristically as a function of the interatomic distance.
In the limit of full separation, the valence orbitals collapse into a single atomic-like energy level.
As the atoms approach one another, this level splits, and the energy separation between the resulting states increases with further reduction of the distance.
While the higher-energy antibonding state typically increases monotonically in energy as distance decreases, the lower-energy bonding state exhibits a distinct energy minimum that determines the equilibrium bond length.
By examining the influence of the Cr--Cr interatomic distance in \ce{CeCrB4} on the behavior of the Cr~$3d_{z^2}$ states, we expect to observe signatures of classical molecular-like states.

%
In our study, the intra-dimer $d_{\text{Cr--Cr}}$ distance was systematically varied by tuning the lattice parameters $a$ and $b$, while maintaining a constant $a/b$ ratio and keeping the lattice parameter $c$ fixed, see Fig.~\ref{fig:crystal_structure}(a).
While the equilibrium distance in \ce{CeCrB4} is $2.40$~\AA{}, we investigated a range of separations from $1.86$ to $3.48$~\AA{}.
To simplify the physical interpretation and rule out competing electronic effects, these calculations were performed neglecting both spin polarization and spin-orbit coupling.

%
The effect of the interatomic distance on the electronic structure of the dimer, illustrated in Fig.~\ref{fig:3dz2_dos}, provides evidence supporting the molecular nature of these localized states.
In the limit of large separations ($d_{\text{Cr--Cr}} \sim 3.48$~\AA{}), a single Cr~$3d_{z^2}$ peak dominates the density of states. 
As the dimer atoms approach one another, this atomic-like level splits into two well-localized features.
At the equilibrium distance of $2.40$~\AA{}, these states begin to experience noticeable broadening due to increased interaction with the surrounding crystalline environment, which ultimately leads to the complete delocalization and blurring of the molecular features below $d_{\text{Cr--Cr}} \sim 2.0$~\AA{}.

%
This distance-induced tuning of the molecular-like states represents a crucial degree of freedom in the \ce{MTB4} family, which in practice can be controlled experimentally via chemical substitution (chemical pressure), hydrostatic pressure, or substrate-induced epitaxial strain in thin-film heterostructures.
Although Fig.~\ref{fig:3dz2_dos} displays the nonmagnetic electronic spectrum, the consideration of magnetism in certain \ce{MTB4} compounds leads to an additional spin-splitting of these $3d_{z^2}$ states, as demonstrated in Fig.~\ref{fig:dos_cecrb4}.
Consequently, the intra-dimer distance modulates the magnetic exchange interactions and may govern potential magnetic phase transitions.
Furthermore, an external magnetic field could drive significant changes in the electronic transport properties of the material, particularly if the field-induced spin splitting forces these localized states to cross the Fermi level.

\subsection{Tuning of molecular-like states in \ce{CeTB4} via transition metal selection}

In \ce{MTB4} systems, an additional degree of freedom governing the energetic position of the localized states of \ce{T}--\ce{T} dimer is the atomic number of the transition metal \ce{T}.
For the \ce{YTB4} and \ce{LuTB4} series containing $3d$ transition metals with successive atomic numbers (V, Cr, Mn, Fe, Co, and Ni), Zhang~\textit{et al.} demonstrated that the localized molecular-like states shift systematically deeper below the Fermi level as the atomic number increases~\cite{zhang_unveiling_2024}.
In Fig.~\ref{fig:local_dos_cetb4}(a), we present our computed \ce{T}~$3d_{z^2}$ DOS for three consecutive $3d$ transition metals within the \ce{CeTB4} family.
In line with the findings of Zhang~\textit{et al.}, a clear rigid-band-like downward shift in the position of the $3d_{z^2}$ states is observed with increasing atomic number of \ce{T}.
From a materials design perspective, this chemical trend offers a robust pathway for tailoring the position of localized states through compositional engineering.

%
Although current and recent works have predominantly focused on \ce{MTB4} borides hosting $3d$ transition metals, an intriguing prospect lies in extending these concepts to isostructural compounds containing $4d$ and $5d$ elements~\cite{sobczak_magnetic_1979,akopov_investigation_2018,benndorf_11b_2019,wei_first-principles_2025}.
For instance, \ce{YTB4} borides with \ce{T}$ = \text{Cr, Mo, W}$ have recently attracted attention as efficient electrocatalysts for the hydrogen evolution reaction~\cite{hossain_probing_2026}, where Cr, Mo, and W represent the $3d$, $4d$, and $5d$ members of the same periodic group, respectively.
Our results presented in Fig.~\ref{fig:local_dos_cetb4}(b) clearly suggest the persistence of molecular-like states in systems with heavier $4d$ and $5d$ congeners, thereby opening up further avenues for property manipulation in \ce{MTB4}-type materials.

\section{Summary and Conclusions}

Using density functional theory within the GGA+$U$ framework, we investigated the electronic structure, charge transfer mechanisms, and ground-state magnetic configurations of the intermetallic compound \ce{CeCrB4}.
The obtained results were contextualized with respect to previous studies on both this specific compound and the broader \ce{MTB4} material family.
We demonstrated that the layered crystal structure of \ce{CeCrB4} is complemented by a distinct charge polarization of the atomic sheets, where the Ce/Cr layers exhibit donor behavior and the B layers act as acceptors.
The occupation of Ce~$4f$ orbital is found to be 0.66, which qualitatively indicates the mixed-valence state of Ce in this system, as previously revealed by experimental spectroscopy.
Furthermore, our calculations indicated that the ferromagnetic ground state of \ce{CeCrB4} is only a fraction of a millielectronvolt more stable than the competing antiferromagnetic configuration, providing an explanation as to why no magnetic phase transition was experimentally observed down to 1.8~K in earlier measurements.

An intriguing feature of the crystal structure of \ce{CeCrB4} is the presence of Cr--Cr dimers, which leads to the formation of molecular-like states within the electronic spectrum.
In this work, we systematically evaluated the energetic position of these localized Cr~$3d_{z^2}$ states as a function of the intra-dimer interatomic distance.
The observed evolution correlates well with the typical bonding and antibonding level splitting characteristic of classical homonuclear diatomic molecules.
Furthermore, calculations incorporating spin polarization for these Cr~$3d_{z^2}$ molecular-like states allow for a detailed microscopic interpretation of the two main classes of magnetic \ce{MTB4} systems: conventional magnets and quantum spin dimers.
Extending these calculations to other members of the \ce{CeTB4} family suggests that such molecular-like states persist in systems hosting heavier transition metals from the $4d$ and $5d$ series as well.
Consequently, by manipulating the chemical composition, hydrostatic pressure, or, in the case of thin-film heterostructures, the epitaxial substrate strain, it is possible to tune the localized states residing in the immediate vicinity of the Fermi level, thereby offering a robust pathway to control both the electrical and magnetic properties of materials within the \ce{MTB4} family.

\section*{Acknowledgements}
We gratefully acknowledge financial support from the National Science Center Poland under decision DEC-2021/41/B/ST5/02894 (OPUS 21).
We thank Paweł Leśniak and Daniel Depcik for compiling the scientific software and administering the computational cluster at the Institute of Molecular Physics, Polish Academy of Sciences.
\end{sloppypar}

\bibliography{cecrb4.bib}

\end{document}